\documentclass[twocolumn,twoside]{article}
\usepackage{graphicx}
\usepackage{hyperref}

\begin{document}

\thispagestyle{plain}
\markboth{\uppercase{\rm Space-Time Description \hspace{57mm}}}
{\hspace{74mm} \uppercase{\rm Pavlov}}

\twocolumn[
\begin{center}
{\LARGE \bf Space-Time Description of Scalar Particle Creation

\vspace{10pt}
by a Homogeneous Isotropic Gravitational Field}
\vspace{19pt}

{\Large \bf {Yu. V. Pavlov}${}^*$} \vspace{9pt}

{\it A. Friedmann Laboratory for Theoretical Physics,
St.\,Petersburg, Russia, and}\\
{\it Institute of Mechanical Engineering, Russian Acad. Sci., 61 Bolshoy pr.,
St. Petersburg 199178, Russia}
\end{center}
\vspace{5pt}
 {\bf Abstract.}
We give the generalization of the method of the space-time description
of particle creation by a gravitational field for a scalar field
with nonconformal coupling to the curvature.
    The space-time correlation function is obtained for a created pair
of the quasi-particles, corresponding to a diagonal form of the instantaneous
Hamiltonian.
The case of an adiabatic change of the metric of homogeneous isotropic
space is analyzed.
We show that the created pairs of quasi-particles in de Sitter space
should be interpreted as pairs of virtual particles.

\vspace{11pt}
{PACS number:} 04.62.+v, 03.70.+k
%% 04.62.+v Quantum fields in curved spacetime
%% 03.70.+k Theory of quantized fields
\vspace{27pt}
]

%%%%******************************************************************
{\centering  \section{\large  \uppercase{Introduction}}}

\footnotetext[1]{E-mail: \, {\tt yuri.pavlov@mail.ru}}

    Quantum field theory in curved space-time is nowadays a sufficiently
deeply elaborated area of theoretical physics
(see the monographs~\cite{GMM,BD}) with important applications in
cosmology and astrophysics.
In particular, creation of particles with GUT-scale masses by
the gravitational field of the early Universe may be used for an explanation
of the observed visible and dark matter density~\cite{GrPv}.

    In the description of particle creation by the gravitational field,
a widely used method is that of instantaneous Hamiltonian
diagonalization~\cite{GMM} suggested by
A.\,A. Grib and S.\,G. Mamayev~\cite{GribMamayev69}.
    A~detailed study of the created pair formation process was performed
in~\cite{MT} for the case of a scalar field conformally coupled to
the curvature.
    The method of a space-time correlation function suggested in~\cite{MT}
made it possible to distinguish real created particles from virtual ones,
to reveal the role of horizons in particle creation etc.

    In quantum field theory in curved space-time, one frequently considers
scalar field nonconformally coupled to gravity, in particular, the minimally
coupled one.
In such cases, the contributions related to nonconformal coupling may be
dominant both in the particle creation effect~\cite{BMR98} and in the vacuum
averages values of the stress-energy tensor (see, e.g.,~\cite{BLMPv}).
Such a coupling with the curvature also takes place in the case of massive
vector mesons (the longitudinal components~\cite{GMM}).
    Conformal invariance is lacking in the graviton equations
as well~\cite{GrishchukY80}.
Therefore, it appears to be necessary to generalize the method of studying
the particle creation process suggested in~\cite{MT} to the nonconformal case.
    Moreover, some authors~\cite{MTrunov03} consider the absence of such
a generalization as an argument in favor of choosing only the conformal
coupling in the wave equation for a scalar field.

    The present paper suggests a generalization of the space-time description
of particle creation~\cite{MT} to the case of nonconformal coupling.
    In \hyperref[2sec]{Section~2}, we perform quantization of a general-type
nonconformally coupled scalar field in homogeneous isotropic
space.
    In \hyperref[3sec]{Section~3}, we carry out diagonalization of the
generalized Hamiltonian built in~\cite{Pv}, which, in the nonconformal case,
allows one to solve the well-known problem of an infinite density
of quasiparticles created~\cite{Fulling79}.
    In \hyperref[4sec]{Section~4}, we build a space-time correlation function
for quasiparticles corresponding to the diagonal form of the
instantaneous Hamiltonian and study the case of an adiabatically
changing metric.
    In \hyperref[5sec]{Section~5}, we consider particle creation and
the space-time correlation function in de Sitter space.
    The \hyperref[6sec]{conclusion} briefly sums up the results of the paper.

    We use the system of units in which $\hbar = c \!=\! 1$.\
    The signs of the curvature tensor and the Ricci tensor are chosen in such
a way that \ $ R_{ik} = R^{\, l}_{\ ilk} $\,,
    $$ R^{\, i}_{\ jkl} = \partial_l \, \Gamma^{\, i}_{\, jk} -
\partial_k \, \Gamma^{\, i}_{\, jl} +
\Gamma^{\, i}_{\, nl} \Gamma^{\, n}_{\, jk} -
\Gamma^{\, i}_{\, nk} \Gamma^{\, n}_{\, jl} \,,
$$
where $\Gamma^{\, i}_{\, jk}$ are the Christoffel symbols.

\newpage    %\vspace{20pt}
%%%%*****************************************************************
{\centering \section{\large  \uppercase{Scalar field in curved space}}
\label{2sec}
}

    Consider a complex scalar field $\varphi(x)$ with mass $m$,
the Lagrangian
    \begin{equation}
L(x)=\sqrt{|g|} \left[\, g^{ik}\partial_i\varphi^*\partial_k\varphi -
(m^2 + V_{\!g})\, \varphi^* \varphi \, \right],
\label{Lag}
\end{equation}
     and the corresponding equation of motion
\begin{equation}
( \nabla^i \nabla_{\! i} + V_{\!g} + m^2 )\, \varphi(x)=0 \,,
\label{Eqm}
\end{equation}
    where ${\nabla}_{\! i}$ are covariant derivatives in
$N$-dimensional space-time with the metric $g_{ik}$,\
$ g\!=\!{\rm det}(g_{ik})$,  $\ V_{\!g}$ is a function of invariant
combinations of the metric tensor $g_{ik}$ and its partial derivatives.

    Eq.~(\ref{Eqm}) is conformally invariant if $m=0$ and
$V_{\!g}=\xi_c R $, where $R$ is the scalar curvature and
$\xi_c = (N-2)/\,[\,4\,(N-1)] $ (conformal coupling).
    The case $V_{\!g}=0$ corresponds to minimal coupling.
    An arbitrary $ V_{\!g} $ leads to the advent of third- and higher-order
derivatives of the metric in the metric stress-energy tensor of the scalar
field and consequently in the Einstein equations.

    It is well known that additional terms with higher-order derivatives
appearing in equations lead to radical changes in the theory even if the
coefficients of these terms are small.

    If one requires that the metric stress-energy tensor should not contain
derivatives of the metric of orders higher than two, then the following
function is admissible as $ V_{\!g} $:
    \begin{equation}
V_{\!g} = \xi R + \zeta R_{GB}^{\,2} \,,
\label{V}
\end{equation}
    where
    \begin{equation}
R_{GB}^{\,2} \stackrel{\rm def}{=}
R_{lmpq} R^{\,lmpq} - 4 R_{lm} R^{\,lm} + R^2
\label{RGB}
\end{equation}
    (the Gauss-Bonnet coupling~\cite{Pv4}).

    Let us note that for $N=4$, with constant $\varphi(x)$, the contribution
to the metric stress-energy tensor from the term with $R_{GB}^{\,2}$ is absent
because the corresponding variation derivative vanishes~\cite{Lanczos38}.
    But for a variable $\varphi(x)$, a contribution from such terms could be
taken into account if the constant $\zeta $ with the dimension (mass)$^{-2}$
is nonzero.

    Accounting for a possible coupling between a scalar field and
the Gauss-Bonnet invariant $R_{GB}^{\,2}$ may play an important role in
the early Universe; effects from a nonzero value of the parameter $ \zeta $
in scalar field equations may appear in black-hole radiation, may affect
the parameters of the so-called boson stars etc.
The question of the values of the parameters $ \zeta $ and $ \xi $
are ultimately related to the area of the experiment.

    Furthermore, without specifying $V_{\!g}$, let us consider an
$N$-dimensional homogeneous isotropic space-time, choosing the metric
in the form
    \begin{equation}
ds^2=g_{ik}dx^i dx^k = a^2(\eta)\,(d{\eta}^2 - d l^2) \,,
\label{gik}
\end{equation}
    where $d l^2=\gamma_{\alpha \beta} d x^\alpha d x^\beta $ is the metric
of an $(N-1)$-dimensional space of constant curvature $K=0, \pm 1 $.

    The complete set of solutions to Eq.~(\ref{Eqm}) in the metric~(\ref{gik})
may be found in the form
    \begin{equation}
\varphi(x) = \frac{\tilde{\varphi}(x)}{a^{(N-2)/2} (\eta)} \,
= a^{-(N-2)/2} (\eta)\, g_\lambda (\eta) \Phi_J ({\bf x})\,,
\label{fgf}
\end{equation}
    where
    \begin{equation}
g_\lambda''(\eta)+\Omega^2(\eta)\,g_\lambda(\eta)=0 \,,
\label{gdd}
\end{equation}
       \begin{equation}
\Omega^2(\eta)=(m^2 + V_{\! g} - \xi_c R) a^2 +\lambda^2 ,
\label{Ome}
\end{equation}
     \begin{equation}
\Delta_{N-1}\,\Phi_J ({\bf x}) = - \Biggl( \lambda^2 -
\biggl( \frac{N-2}{2} \biggr)^2 K \Biggr) \Phi_J  ({\bf x})\,,
\label{DFlF}
\end{equation}
    the prime denotes a derivative with respect to the conformal time $\eta$,
and $J$ is the set of indices (quantum numbers) numbering the eigenfunctions
of the Laplace-Beltrami operator $\Delta_{N-1}$
in ($N\!-\! 1$)-dimensional space.

    According to the Hamiltonian diagonalization method~\cite{GMM}
(see the case of an arbitrary function $V_{\!g}$ in~\cite{PvIJA}),
the functions $g_\lambda(\eta)$ should obey the following initial conditions:
    \begin{equation}
g_\lambda'(\eta_0)=i\, \Omega(\eta_0)\, g_\lambda(\eta_0) \,, \ \
\ |g_\lambda(\eta_0)|= \Omega^{-1/2}(\eta_0)\,.
\label{icg}
\end{equation}

    To perform quantization, let us expand the field $ \varphi(x) $
in the complete set of solutions~(\ref{fgf})
    \begin{equation}
\varphi(x)=\int \! d\mu(J)\,\biggl[ \varphi^{(+)}_J \,a^{(+)}_J +
\varphi^{(-)}_J \, a^{(-)}_J \,\biggr],
\label{fff}
\end{equation}
    where $d\mu(J)$ is a measure on the set of quantum numbers,
    \begin{equation}
\varphi^{(+)}_J (x)\!=\!\frac{g_\lambda(\eta)\,\Phi^*_J({\bf x})}
{\sqrt{2}\, a^{(N-2)/2}(\eta)}, \ \ \varphi^{(-)}_J(x) =
\bigl(\varphi_J^{(+)}(x)\! \bigr)^* \!,
\label{fpm}
\end{equation}
    and require that the standard commutation relations hold for
$a^{(\pm)}_J\!$ and $\stackrel{*}{a}\!{\!}^{(\pm)}_J$.

    Let us build the Hamiltonian as the canonical one for the variables
$\tilde{\varphi}(x)$ and $\tilde{\varphi}^*(x)$, for which the equation
of motion does not contain their first-order derivatives with respect to
the time $\eta$~\cite{PvIJA}.
    Recall that the equations of motion do not change after adding a full
divergence $\partial J^i / \partial x^i$ to the Lagrangian density $L(x)$.

    Let us choose, in the coordinate system $(\eta, {\bf x})$, the vector
$$
(J^i)=(\,\sqrt{\gamma} c \tilde{\varphi}{}^* \tilde{\varphi} (N-2)/2, 0,
\ldots , 0\,), $$
    where $\gamma={\rm det}(\gamma_{\alpha \beta})$, $c=a'/a$.
    Then, using the Lagrangian density
$ L^{\Delta}(x)=L(x) + \partial J^i / \partial x^i $,
we obtain for the momenta canonically conjugate to $\tilde{\varphi}$ and
$\tilde{\varphi}^*$:
    \begin{equation}
\pi \equiv \frac{\partial L^{\Delta}}{\partial \tilde{\varphi}'}=
\sqrt{\gamma}\, \tilde{\varphi}^*{}' \ , \ \ \
\pi_* \equiv \frac{\partial L^{\Delta}}{\partial
\tilde{\varphi}^{* \prime}} = \sqrt{\gamma}\, \tilde{\varphi}',
\label{imp}
\end{equation}
    respectively.
    Integrating the Hamiltonian density $h(x)=\tilde{\varphi}{}' \pi +
\tilde{\varphi}^{* \prime} \pi_* - L^{\Delta}(x) $
over the hypersurface $\Sigma$: $\eta = {\rm const} $,
we obtain the following expression for the canonical Hamiltonian:
    \begin{eqnarray}
H(\eta) \!=\!
\int_\Sigma d^{N-1}x \, \sqrt{\gamma} \, \biggl\{
\tilde{\varphi}^{* \prime} \tilde{\varphi}^\prime
+ \gamma^{\alpha \beta} \partial_\alpha\tilde{\varphi}^*
\partial_\beta \tilde{\varphi} {} +
\nonumber                 \\
\biggl[ \left(m^2 \! + V_{\!g} \right) a^2 - \frac{N\!-2}{4} \left(2c'+
(N \!-2)c^2\right) \biggr] \tilde{\varphi}^* \tilde{\varphi} \biggr\}
\label{hx}
\end{eqnarray}
    (see a justification of such a choice of the Hamiltonian
in~\cite{Pv,PvIJA} and in Section~3).

    The Hamiltonian~(\ref{hx}) may be written in terms of the operators
$ a_J^{(\pm)} $ and $ \stackrel{*}{a}\!{\!}_J^{(\pm)} $ in the following way:
    \begin{eqnarray}
H(\eta)\! &=& \! \int \! d\mu(J) \, \biggl[ E_J(\eta)
\left( \stackrel{*}{a}\!{\!}^{(+)}_J a^{(-)}_J +
\stackrel{*}{a}\!{\!}^{(-)}_{\bar{J}} a^{(+)}_{\bar{J}} \right) +
  \nonumber \\
&+& F_J(\eta)  \stackrel{*}{a}\!{\!}^{(+)}_J a^{(+)}_{\bar{J}} +
F^*_J(\eta) \stackrel{*}{a}\!{\!}^{(-)}_{\bar{J}} a^{(-)}_J  \biggr],
\label{H}
\end{eqnarray}
\vspace*{-2mm}
    where
\vspace{-1mm}
\begin{equation}
E_J=\frac{|g_\lambda'|^2+ \Omega^2 |g_\lambda|^2}{2},
\ \ \ F_J=\frac{\vartheta_{\!J}}{2} \bigl[ g_\lambda'{}^{\! 2} +
\Omega^2  g_\lambda^{\, 2} \bigr],
\label{EJFJ}
\end{equation}
and we have chosen such eigenfunctions $\Phi_{\!J}({\bf x})$ that,
for arbitrary $J$, there is such $\bar{J}$ that
$
\Phi_{\!J}^*({\bf x}) = \vartheta_{\!J} \Phi_{\!\bar{J}}({\bf x}),
\ |\vartheta_J|=1  \,,
$
($\ \bar{\!\!\bar{J}} = J$, \ $\vartheta_{\!\bar{J}}=\vartheta_{\!J}$).
    Such a choice is possible due to completeness and orthonormality of
the set $\Phi_{\!J}({\bf x})$.

    In spherical coordinates of a homogeneous isotropic space,
if $J=\{\lambda, l, \ldots, m\}$, we have
$\bar{J}=\{\lambda, l, \ldots, -m\}, \ \vartheta_{\!J}=(-1)^m $
(see~\cite{GMM}).

\vspace{20pt}
%%%%*****************************************************************
{\centering \section{\large  \uppercase{Hamiltonian diagonalization}}
\label{3sec}
}

    The Hamiltonian (\ref{H}) will be diagonal at the time instant~$\eta_0$
with respect to the operators
$ a_J^{(\pm)}$, $ \stackrel{*}{a}\!{\!}_J^{(\pm)} $,
under the conditions~(\ref{icg}).
    Diagonalization of the Hamiltonian at an arbitrary time instant~$\eta $
is carried out in terms of the operators
$ \stackrel{*}{b}\!{\!}^{(\pm)}_J(\eta) $ and $ b^{(\pm)}_J(\eta) $,
connected with
$ \stackrel{*}{a}\!{\!}^{(\pm)}_J $,  $ a^{(\pm)}_J $
by time-dependent Bogoliubov transformations:
    \begin{equation}
\left\{  \begin{array}{c}
a_J^{(-)}=\alpha^*_J(\eta) \,b^{(-)}_J(\eta)-
\beta_J(\eta) \vartheta_{\!J} \, b^{(+)}_{\bar{J}}(\eta) \,,  \\[3mm]
\stackrel{*}{a}\!{\!}_J^{(-)}=\alpha^*_J(\eta)
\stackrel{*}{b}\!{\!}^{(-)}_J\!(\eta)-
\beta_J(\eta) \vartheta_{\!J}
\stackrel{*}{b}\!{\!}^{(+)}_{\bar{J}}\!(\eta) \,,
\end{array} \right.
\label{db}
\end{equation}

\noindent
    where the functions $\alpha_J(\eta)=\alpha_{\bar{J}}(\eta)$ and
$ \beta_J(\eta)=\beta_{\bar{J}}(\eta) $ satisfy the initial conditions
$|\alpha_J(\eta_0)| = 1$,\ $\beta_J(\eta_0) \!=\! 0 $ and the identity
$|\alpha_J(\eta)|^2 \!- |\beta_J(\eta)|^2 \!=\! 1$. \ \
    Substituting~(\ref{db}) and the conjugate expressions to~(\ref{H}),
one can obtain an expression for the Hamiltonian having the same form~(\ref{H})
but with the replacement
$ \stackrel{*}{a}\!{\!}^{(\pm)}_J $,  $ a_J^{(\pm)} \to $
$ \stackrel{*}{b}\!{\!}^{(\pm)}_J $,  $ b_J^{(\pm)}$ and
    \begin{equation}
E_J \to {}_bE_J \!=\! E_J(|\alpha_J|^2 \!+|\beta_J|^2) \!-
2 {\rm Re \,} (F_J \alpha_J \beta_J^* \vartheta_{\!J}^*),
\label{bEJ}
\end{equation}
    \begin{equation}
F_J \to {}_bF_J=-2\alpha_J\beta_J \vartheta_{\!J}E_J + \alpha_J^2 F_J
+\beta_J^2 \vartheta_{\!J}^2 F_J^* .
\label{bFJ}
\end{equation}
    From the diagonality requirement for the Hamiltonian with respect to
the operators
$ {\stackrel{*}{b}\!{\!}^{(\pm)}_J(\eta)} $ and  $ b^{(\pm)}_J(\eta) $
at the time~$\eta$, i.e., $ {}_bF_J(\eta)=0 $,
(for $ \Omega^2(\eta)>0 $), it follows:
    \begin{equation}
\alpha_J=i \chi_J^{\phantom{*}} \,
\frac{g_\lambda^{*\, \prime} \! - i \Omega\, g_\lambda^*}{2 \sqrt{\Omega}} \,,
\ \ \
\beta_J = i \chi_{J}^{\phantom{*}} \,
\frac{g_\lambda^{\, \prime} \! - i \Omega\, g_{\lambda}}{2 \sqrt{\Omega}},
\label{abxi}
\end{equation}
    where $ \chi_J^{\phantom{*}} = \chi_{\bar{J}} $
is an arbitrary complex function of time with a unit absolute value.
Therefore further we will use the operators
    \begin{equation}
c^{(+)}_J (\eta) \!=\! \chi_J^{\phantom{*}}(\eta) \, b^{(+)}_J \!(\eta),
\ \ \  c^{(-)}_J (\eta) \!=\! \chi_J^*(\eta) \, b^{(-)}_J \!(\eta),
\label{cb}
\end{equation}
    which, due to~(\ref{db}) and (\ref{abxi}), do not depend on the specific
choice of the functions $\chi_J^{\phantom{*}}(\eta)$.\ \
    The operators
$ \stackrel{*}{c}\!{\!}^{(\pm)}_J(\eta) $ and $ c^{(\pm)}_J(\eta) $
obey the same commutation relations as
$ \stackrel{*}{a}\!{\!}^{(\pm)}_J $ and  $ a^{(\pm)}_J $.

    Substitution of~(\ref{cb}), taking into account~(\ref{db}) and their
conjugate expressions, into~(\ref{H}) leads to the following expression
for the Hamiltonian:
     \begin{equation}
H(\eta) =\int \! d\mu(J) \,\Omega(\eta)
\left( \stackrel{\,*}{c}{\!}^{(+)}_J c^{(-)}_J +
\stackrel{\,*}{c}{\!}^{(-)}_{J} c^{(+)}_{J} \right).
\label{Hcc}
\end{equation}
    Thus the energy of quasiparticles corresponding to the diagonal form of
the Hamiltonian~(\ref{H}) is equal to the oscillator frequency~$\Omega(\eta)$
(unlike the Hamiltonian built from the metric stress-energy tensor of a
nonconformal scalar field~\cite{BMR98}).

    Using~(\ref{gdd}), (\ref{db}), (\ref{abxi}), and (\ref{Hcc}),
one can verify that the operators $c^{(\pm)}_{J}(\eta)$
obey Heisenberg's equations of motion:
    \begin{equation}
\frac{d c^{(\pm)}_J}{d \eta} = i \left[ H(\eta), c^{(\pm)}_J \right]
+ \frac{\Omega^\prime}{2 \Omega} \,
\vartheta^{(\mp1)}_{\!J} c^{(\mp)}_{\bar{J}} .
\label{dc}
\end{equation}
    The second term in the right-hand side of~(\ref{dc}) is connected with
re-definition of the particle notion at each time instant.

    An expansion of the field operator $\tilde{\varphi}(x)$
in the operators $c^{(\pm)}_J (\eta)$:
    \begin{equation}
\tilde{\varphi}(x)=\! \int \! \frac{d\mu(J)}{\sqrt{2 \Omega}}
\left[ \Phi_{\!J}^*({\bf x}) \, c^{(+)}_J(\eta) +
\Phi_{\!\bar{J}}({\bf x}) \, c^{(-)}_{\bar{J}}(\eta) \right]
\label{ffc0}
\end{equation}
    follows from~(\ref{fff}), (\ref{fpm}), (\ref{db}), and (\ref{abxi}).
    The equations of motion hold for each mode in~(\ref{ffc0}) separately.

    Consider the question of particle creation in a nonstationary metric.
    We suppose that the quantized scalar field is in the state $|0\rangle$,
annihilated by the operators
$ a^{(-)}_J $,\,  $ \stackrel{*}{a}\!{\!}^{(-)}_J $,
i.e., in the vacuum state for the instant~$\eta_0$.
    At the time $\eta$, the vacuum state is the state $|0_\eta\rangle$,
defined by the equalities
    \begin{equation}
c^{(-)}_J(\eta)\, |0_\eta\rangle = \
\stackrel{*}{c}{\!}^{(-)}_J(\eta)\, |0_\eta\rangle=0\,.
\label{vac}
\end{equation}
    The state $|0\rangle$ contains, at the time instant $\eta$, \
$|\beta_J(\eta)|^2$ pairs of particles and antiparticles corresponding
to the operators $c^{(\pm)}_J(\eta)$ (see~\cite{GMM}).
    The density of the created particle pairs may be calculated
(for the quasi-Euclidean metric with $K=0$) by the formula
    \begin{equation}
n(\eta) = \frac{B_N}{2 a^{N-1}} \int \limits_0^\infty \! S_\lambda(\eta)\,
\lambda^{N-2}\, d \lambda,
\label{nN}
\end{equation}
where $B_N=\left[2^{N\!-3} \pi^{(N \!-1)/2} \Gamma((N\!-1)/2) \right]^{-1}\!,$
\ $\Gamma(z)$ is the gamma function,
$ S_\lambda(\eta) = |\beta_\lambda (\eta)|^2 $
(in a homogeneous isotropic space, $ |\beta_J| \equiv |\beta_\lambda| $).
    For $N=4$ and $K=0,-1$, for the number density of the particle pairs
created, the following formula is valid (see~\cite{GMM}):
    \begin{equation}
n(\eta) = \frac{1}{2 \pi^2 a^{3}} \int \limits_0^\infty \! S_\lambda(\eta)\,
\lambda^{2}\, d \lambda,
\label{nN4Km1}
\end{equation}
    For $K=1$ (spherical space), the set of eigenfunctions of
the Laplace-Beltrami operator $\Delta_{N-1}$ is discrete,
and the formula for the number density of the created particle pairs
$N=4$ has the form (see~\cite{GMM})
    \begin{equation}
n(\eta) = \frac{1}{2 \pi^2 a^{3}}
\sum \limits_{\lambda=1}^\infty  S_\lambda(\eta) \lambda^{2}\,.
\label{nN4Kp1}
\end{equation}
    Using (\ref{abxi}) and that the function
$$
g_\lambda(\eta) g^*_\lambda{}^\prime (\eta) -
g_\lambda^\prime(\eta) g^*_\lambda{} (\eta)
$$
is a first integral of the equation~(\ref{gdd}), equal to $-2i$
according to the initial conditions~(\ref{icg}), we obtain:
    \begin{equation}
S_\lambda(\eta) = \frac{1}{4 \Omega} \left( |g'_\lambda|^2 +
\Omega^2 |g_\lambda|^2 \right) - \frac{1}{2} \,.
\label{Sgg}
\end{equation}

    As it was shown in~\cite{Pv}, $S_\lambda(\eta) \sim \lambda^{-6}$
as $\lambda \to \infty$.
    Therefore, in four-dimensional space-time, the number density of particles
created, defined by the Hamiltonian~(\ref{hx}) diagonalization method, is
finite in the nonconformal case as well.
    Let us note that a divergent expression for the number density of created
nonconformal scalar particles, obtained with another choice of the Hamiltonian
in Ref.~\cite{Fulling79}, has been one of the reasons for a criticism of the
Hamiltonian diagonalization method as a whole in the well-known book~\cite{BD}.

\vspace{20pt}
%%%%*****************************************************************
{\centering \section{\large  \uppercase{The space-time correlation
function}}  \label{4sec}
}

    To study the space-time characteristics of the created quasiparticles,
we apply the approach suggested in~\cite{MT}.
    We use the notion of a particle's localized state introduced by
Newton and Wigner~\cite{NewtonWigner49}.
    By analogy with the Newton-Wigner operator for a free field
(see, e.g., \cite{Schweber}), we introduce creation operators
of a localized state of a particle and an antiparticle:
    \begin{eqnarray}
\stackrel{*}{\varphi}{\!\!}_{\,1}^{(+)}(\eta, {\bf x})=
a^{\textstyle -\frac{N\!-1}{2}}(\eta) \int \!\! d \mu (J) \, \Phi^*_J({\bf x})
\stackrel{*}{c}{\!}_{J}^{(+)}(\eta) , \nonumber  \\
             {\varphi}_{\,1}^{(+)}(\eta, {\bf x})=
a^{\textstyle -\frac{N\!-1}{2}}(\eta) \int \!\! d \mu (J) \, \Phi^*_J({\bf x})
\, {c}{}_{J}^{(+)}(\eta) \,.
\label{phiod}
\end{eqnarray}

    By analogy with the case of a conformal scalar field
considered in~\cite{GMM}, we introduce the operator of particle
number in the volume~$V$
    \begin{equation} \label{NV}
N_V =\! \int \limits_V \! d^{N-1} x \, \sqrt{{}^{(N \!-\! 1)}\!g}\,
\stackrel{*}{\varphi}{\!\!}_{\,1}^{(+)}(\eta, {\bf x})\,
{\varphi}_{\,1}^{(-)}(\eta, {\bf x}),
\end{equation}
    where
$ {}^{(N \!-\! 1)}\!g = {\rm det}\left( {}^{(N \!-\! 1)}g_{\alpha \beta}
\right)$, and
$$
{}^{(N \!- 1)}g_{\alpha \beta} = a^2(\eta) \gamma_{\alpha \beta}
$$
is the induce metric tensor on the hypersurface $\eta={\rm const}$.
    Using (\ref{db}), (\ref{cb}), (\ref{phiod}) and the properties of
the eigenfunctions $ \Phi_J({\bf x}) $ (see, e.g., \cite{GMM}, \S\,9.1),
it can be shown that the expression for the number density
of the particles created
    \begin{equation} \label{nNV}
n = \langle 0 |\, N_V | 0 \rangle \, /\, V
\end{equation}
    obtained with the aid of~(\ref{NV}), reproduces
Eqs.~(\ref{nN})--(\ref{nN4Kp1}).

    As in~\cite{MT}, we consider as a characteristic of the spatial
distribution of the quasiparticle pairs created, the matrix element
    \begin{equation} \label{Rtx}
R_0(\eta, {\bf x},{\bf x'})= \frac{\langle 0_\eta | \, \varphi_{\, 1}^{(-)}
(\eta, {\bf x}) \stackrel{*}{\varphi}{\!\!}_{\,1}^{(-)}(\eta, {\bf x'})
\, | 0 \rangle}
{\langle 0_\eta |\, 0 \rangle} \,,
\end{equation}
    which has the meaning of the probability amplitude that a quasiparticle
created is located at the point~$\bf x$ at the time instant~$\eta$ while
the antiquasiparticle is at the point~$\bf x'$.
    Using (\ref{db}), (\ref{abxi}), (\ref{cb}) and (\ref{phiod}), we obtain:
    \begin{equation} \label{RHI}
R_0(\eta, {\bf x},{\bf x'}) \!=\! \frac{1}{a^{N-1}(\eta)}\!
\int \! \! d \mu (J) \Phi_{\! J}({\bf x})
\Phi^*_{\! J}({\bf x'}) P_{\! \lambda}(\eta),
\end{equation}
    where
\begin{equation} \label{PlG}
P_{\! \lambda}(\eta) = ( i \Omega g_\lambda - g_\lambda^{ \, \prime} ) /
( i \Omega g_\lambda + g_\lambda^{\, \prime}) \,.
\end{equation}
    The function $P_{\! \lambda}(\eta)$ satisfies the following equation and
initial condition:
    \begin{equation} \label{Peq}
P_{\! \lambda}^{\, \prime} + 2 i \Omega P_{\! \lambda} +
\frac{\Omega'}{2 \Omega} (P^2_{\! \lambda} - 1) =0 \,,
\ \ \ \ \ P_{\! \lambda}(\eta_0)=0  \,.
\end{equation}

    Consider the case that the metric is changing adiabatically:
    \begin{equation} \label{AIM}
\frac{1}{\Omega} \biggl| \left( \frac{\Omega'}{\Omega^2} \right)^{\!\! \prime}
\biggr| \ll \left| \frac{\Omega'}{\Omega^2} \right| \ll 1 \,.
\end{equation}
    Furthermore, we denote $M=\sqrt{\Omega^2 -\lambda^2} / a$.
In the general case, $M=M(\eta)$.
For a conformally coupled scalar field, $M=m$.
If $M={\rm const}$, the conditions~(\ref{AIM}) hold if
$\dot{h}(t)/M^2 \ll h/M \ll 1$,
where $h(t)=\dot{a}(t)/a$ is the Hubble parameter.

    If the conditions~(\ref{AIM}) and $a'(\eta_0)=0 $ are satisfied,
an approximate solution to~(\ref{Peq}) has the form
    \begin{equation} \label{PJAd}
P_{\! J}(\eta) \approx -i \Omega' / (4\Omega^2) \,.
\end{equation}

Let us find an expression for the space-time correlation function~(\ref{Rtx})
in the approximation considered in the quasi-Euclidean metric
(i.e., $ d s^2= a^2(\eta) ( d \eta^2 - dx^\alpha d x^\alpha )$).
The eigenfunctions of the Laplace operator $\Delta_{N-1}$ in the coordinates
$ x^\alpha $ are
    \begin{equation} \label{FD}
\Phi_J({\bf x}) = (2 \pi )^{(1-N)/2} \exp (-i \lambda_\alpha x^\alpha ) \,,
\end{equation}
    where $-\infty < \lambda_a < +\infty,
\ \ \sum_\alpha \lambda_\alpha^2 = \lambda^2 $.
    Consequently,
\begin{equation} \label{F2} \hspace{-20pt}
\sum \limits_{\hspace{7mm} J\, (\lambda={\rm const})}\hspace{-7mm}
\Phi_{\! J}^*({\bf x}) \Phi_{\! J}({\bf x'}) =
\Bigl( \frac{\lambda}{2 \pi} \Bigr)^{\!\!\! \frac{N-1}{2 \mathstrut}}
\frac{J_{(N-3)/2}(\lambda \rho)}{\rho^{(N-3)/2}},
\end{equation}
    where $\sum = \int d \! \stackrel{\rightarrow}{\lambda}
 \delta(|\!\stackrel{\rightarrow}{\lambda}\!| -\lambda)$, \
$\rho=|{\bf x} - {\bf x'}|$,  and  $ J_\nu(z)$ are Bessel functions.
Substituting~(\ref{PJAd}), (\ref{F2}) into~(\ref{RHI})
and performing integration, we obtain
    \begin{equation} \label{REu} \hspace{-4pt}
R_0(\eta, {\bf x},{\bf x'}) \!=
\frac{-i (Ma)^{2\, \prime}}{16 \pi^2 a^3} \Bigl( \frac{M}{2 \pi r}
\Bigr)^{\!\!\! \frac{N-4}{2}}\! K_{\frac{N-4}{2}} (M r), \hspace{-2pt}
\end{equation}
where $r=\rho a$ and $ K_\nu(z) $ are MacDonald's functions.

    If $M={\rm const}$, which is the case for a conformally coupled scalar
field and for an arbitrary coupled field in de Sitter space,
from~(\ref{REu}) we obtain
    \begin{equation} \label{REM} \hspace{-3pt}
R_0(t, {\bf x},{\bf x'}) \!=
\frac{-i M^2 h(t)}{8 \pi^2} \Bigl( \frac{M}{2 \pi r}
\Bigr)^{\!\!\! \frac{N-4}{2}}\! K_{\frac{N-4}{2}} (M r).
\end{equation}

    Let us further consider spatial sections with $K= \pm 1$.
For $N=4$, the space-time metric may be written in the form
    \begin{equation} \label{SM}
dl^2= d \chi^2 + f^2(\chi)
\left( d\vartheta^2 + \sin^2 \vartheta \, d \varphi^2 \right),
\end{equation}
where $f(\chi) = \sinh (\chi), \ \chi, \ \sin(\chi) $
for $K=-1, 0, +1$, respectively.
Meanwhile,
    \begin{equation} \label{F2N4}
\sum \limits_{\hspace{7mm} J\, (\lambda={\rm const})}\hspace{-7mm}
\Phi_{\! J}^*({\bf x}) \Phi_{\! J}({\bf x'}) =
\frac{\lambda}{2 \pi^2} \, \frac{\sin \lambda \rho}{f(\rho)}\,,
\end{equation}
where $\rho$ is the geodesic distance between the points
$ {\bf x}$ and  ${\bf x'}$ (see~\cite{GMM}).
    Therefore, from~(\ref{RHI}) and (\ref{PJAd}) we obtain in
the hyperbolic ($K=-1$) case:
    \begin{equation} \label{REuG4}
R_0(\eta, {\bf x},{\bf x'}) =
\frac{-i (Ma)^{2\, \prime}}{16 \pi^2 a^3}
\frac{\rho}{\sinh \rho}\, K_0 (M r).
\end{equation}
   If $|z| \gg 1$, then $K_\nu(z) \sim \sqrt{\pi/(2z)}\ e^{-z}$.
Consequently, if the metric is changing adiabatically, the function $|R_0|$
in the quasi-Euclidean (see~(\ref{REu})) and hyperbolic space-time decreases
exponentially with growing $\rho$ at spacings $r \gg 1/M$,
i.e., exceeding the Compton wavelength.

    In spherical case, the spacing $\rho$ changes in the range
$0 \le \rho \le \pi$.
A substitution of~(\ref{PJAd}) and (\ref{F2N4}) into (\ref{RHI})
leads to
    \begin{equation} \label{REuS4}
R_0(\eta, {\bf x},{\bf x'}) =
\frac{-i (Ma)^{2\, \prime}}{16 \pi^2 a^3 \sin \rho}
\sum \limits_{\lambda=1}^\infty \frac{\lambda \sin \lambda \rho}{
(M^2 a^2 + \lambda^2)^{3/2}} \,.
\end{equation}
    The sum in the right-hand side of~(\ref{REuS4}), after a chain of
transformations, may be presented in the form
    \begin{eqnarray}
\sum \limits_{n=0}^\infty(\rho \!+\! 2 \pi n)\,
K_0 \left( M a ( \rho \!+\! 2 \pi n) \right) -    \nonumber \\
- \, \sum \limits_{n=1}^\infty(2 \pi n \!-\! \rho)\,
K_0 \left( M a (2 \pi n \!-\! \rho) \right).
\label{IP}
\end{eqnarray}
    From the asymptotic properties of the function $K_0(z)$ it follows that,
for $\rho \ll 1 $ and $M a \gg 1$ (the distance between particles od a pair
and a particle's Compton wavelength are much smaller than the curvature radius
of space), in this representation one could retain only the term
$ \rho \, K_0 ( M a \rho) $, and therefore Eq.~(\ref{REuS4}) takes the form
    \begin{equation} \label{RAS4}
R_0(\eta, {\bf x},{\bf x'}) =
\frac{-i (Ma)^{2\, \prime}}{16 \pi^2 a^3 }
\frac{\rho}{\sin \rho } \, K_0 ( M r) \,.
\end{equation}
    Therefore, in the spherical case $|R_0|$ also decreases exponentially
with growing $\rho$ at distances $r \gg 1/M$,
exceeding the Compton wavelength.

    Thus if the metric of a homogeneous isotropic state is changing
adiabatically, the space-time correlation function
$ R_0(\eta, {\bf x},{\bf x'}) $ is exponentially small for $r \gg 1/M$.
This indicates that the corresponding quasiparticles are virtual pairs with
the characteristic correlation length equal to $1/M$.
Real particle creation is exponentially small and does not manifest
itself in perturbation theory.

\vspace{20pt}
%%%%*****************************************************************
{\centering
\section{\large \uppercase{Particle creation \hspace{3cm} in de Sitter space}}
\label{5sec}
}

Let us consider de Sitter space and take the metric in the form~(\ref{gik})
with $K=0$ and
    \begin{equation} \label{dS1}
a=a_1 e^{Ht}=-\frac{1}{H\eta}\,,
\end{equation}
$t \in (-\infty,+\infty) \ \Leftrightarrow \ \eta \in (-\infty, \, 0)$.

    Solutions to Eq.~(\ref{gdd}) with $V_{\!g}=\xi R$ and
the conditions~(\ref{icg}) for $\eta_0 \to -\infty$ have the form
    \begin{equation} \label{dSg}
g_\lambda(\eta)= \sqrt{-\frac{\pi \eta}{2}}
\ e^{{\textstyle \frac{\pi}{2}}\, {\rm Im}\, \nu}
H_\nu^{(2)}(-\lambda \eta)\, e^{i \alpha_0}\,,
\end{equation}
    where $ H_\nu^{(2)}(z) $ is a Hankel function,
$\alpha_0$ is an arbitrary real constant,
    \begin{equation} \label{dSnuM}
\nu=\sqrt{\frac{1}{4} - \left(\frac{M}{H} \right)^2}, \ \ \
M=\sqrt{m^2+(\xi-\xi_c)R},
\end{equation}
    and $R=N(N-1)H^2$.

\begin{figure}[t]
\centering
   \includegraphics[width=79mm]{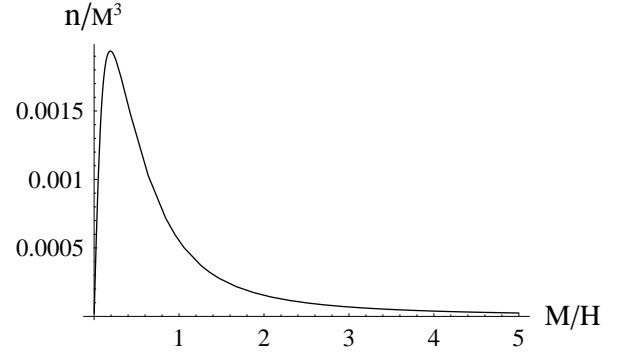}\\
\caption{The number density of quasiparticles created in de Sitter space.}
  \label{dSn}
\end{figure}

    Furthermore, assuming $m^2+(\xi-\xi_c)R > 0$, from
(\ref{nN}), (\ref{Sgg}) and (\ref{dSg}) we obtain
    \begin{equation} \label{dSdens}
n = M^{N-1}\!\cdot F_N \biggl(\frac{M}{H} \biggr),
\end{equation}
i.e., the created particle number density in de Sitter space is
time-independent!
The result of a numerical computation for the function $F_N (M/H)$
in the case $N=4$ is represented in Fig.~\ref{dSn}.
%%%%%%%%%%%%%%%%%%%%

    At $M/H \gg 1 $, the metric is changing adiabatically and, as shown in
Section~4, real particle creation does not occur.

    In the general case, substituting the exact solution~(\ref{dSg})
into~(\ref{RHI}) and (\ref{PlG}) and using~(\ref{F2}), we obtain
    \begin{equation} \label{dSRM}
\frac{R_0(t, {\bf x},{\bf x'})}{M^{N-1}} = f \left( \! Mr, \frac{M}{H} \right),
\end{equation}
where $r=a(t) |{\bf x'} - {\bf x}|$,
i.e., the correlation function, expressed in terms of $r$
(the ``physical'' distance between the quasiparticles in a pair) is
time-independent.

\begin{figure}[t]
\centering
   \includegraphics[width=79mm]{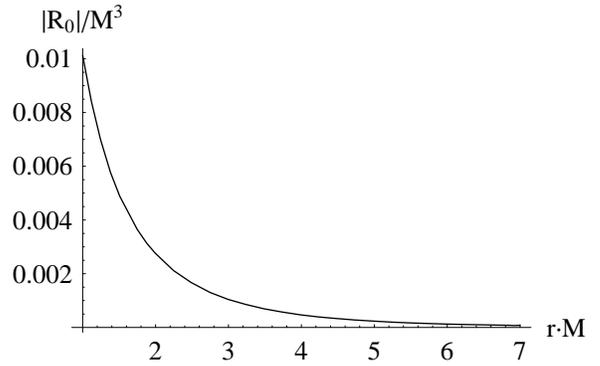}\\
\caption{The correlation function for $M/H=0.01$.}
  \label{dSkMH0.01}
\end{figure}

\begin{figure}[t]
\centering
   \includegraphics[width=79mm]{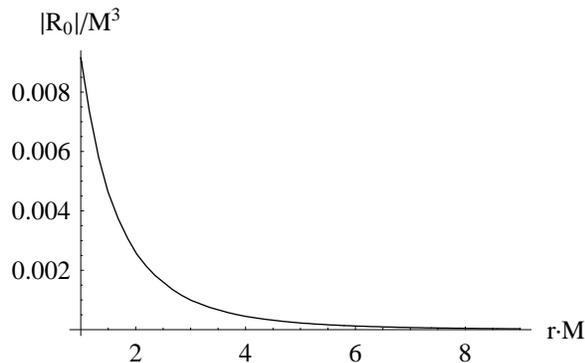}\\
\caption{The correlation function for $M/H=0.2$.}
  \label{dSkMH0.2}
\end{figure}

    Examples of numerical calculations for $N=4$, $M/H=0.01$ and $M/H=0.2$
are given in Figs.~\ref{dSkMH0.01} and \ref{dSkMH0.2}, respectively.
%%%%%%%%%%%%%%%%%%%%
%%
%%%%%%%%%%%%%%%%%%%%

In both cases, the correlation function decreases exponentially at distances
between the quasiparticles exceeding the Compton wavelength $l_C=1/M$.

    Thus real particles creation in de Sitter space does not occur.
The quasiparticle pairs being created, whose density has been calculated and
shown in Fig.~\ref{dSn}, should be interpreted as pairs of virtual particles.

    As has been noticed in~\cite{GMM}, the absence of real particle creation
in de Sitter space is confirmed by the local nature of the vacuum
stress-energy tensor and by vanishing of the imaginary part of
the effective Lagrangian.

\vspace{20pt}
%%%%*****************************************************************
{\centering \section{\large \uppercase{Conclusions}}
\label{6sec}
}

    In this paper, for a scalar field nonconformally coupled to the curvature,
we gave a generalization of the method of space-time description of particle
creation by the gravitational field.
    In a homogeneous isotropic space, we have introduced the creation
operators~(\ref{phiod}) of localized one-particle states and
the operator~(\ref{NV}) of particle number in a specified volume.
    We have obtained the expressions~(\ref{RHI}) and (\ref{PlG}) for
the space-time correlation function~(\ref{Rtx}) of a pair of created
quasiparticles corresponding to a diagonal form of the instantaneous
Hamiltonian.
    We have analyzed the case of adiabatic changes in the metric of
a homogeneous isotropic space.
    The expressions~(\ref{REu}), (\ref{REM}), (\ref{REuG4}) and (\ref{REuS4})
have been obtained for the correlation function of a pair of quasiparticles
created.
    It has been shown that the correlation function exponentially decreases
at spaces exceeding the Compton wavelength, and consequently real particle
creation is suppressed.
    Particle creation in de Sitter space has been considered, and, from,
the properties of the space-time correlation function for a pair of
quasiparticles created, it has been concluded that such a pair should be
interpreted as a pair of virtual particles.

\vspace{19pt}
{\centering \section*{\large \rm \uppercase{Acknowledgments}}}

    The author thanks Prof. A.A.\,Grib for helpful discussions.
The work has been financially supported by RNP Grant 2.1.1.6826.

\vspace{17pt}
%%%%%%%%%%%%%%%%%%%%%%%%%%%%%%%%%%%%%%%%%%%%%%%%%%%%%%%%%%%%%%%%%%%%%%


\begin{thebibliography}{30}
% \itemsep=0pt

\bibitem{GMM}
A.\,A. Grib, S.\,G. Mamayev, and V.\,M. Moste\-panenko,
{\it Vacuum Quantum Effects in Strong Fields\/}
(Energoatomizdat, Moscow, 1988, in Russian;
English translation: \ Friedmann Lab. Publ., St.Petersburg, 1994).

\bibitem{BD}
N.\,D. Birrell and P.\,C.\,W. Davies, {\it Quantum Fields in Curved Space\/}
(Cambridge Univ. Press, Cambridge, 1982).

\bibitem{GrPv}
A.\,A. Grib and Yu.\,V. Pavlov,
{Int. J. Mod. Phys.} D {\bf 11}, 433 (2002);\
{Int. J. Mod. Phys.} A {\bf 17}, 4435 (2002);\
{Grav. \& Cosmol.} {\bf 12}, 159 (2006);\
{Grav. \& Cosmol.} {\bf 14}, 1 (2008).

\bibitem{GribMamayev69}
A.\,A. Grib and S.\,G. Mamayev,
{Yadernaya Fizika} {\bf 10}, 1276 (1969);
{Sov. J. Nucl. Phys.} {\bf 10}, 722 (1970).

\bibitem{MT}
S.\,G. Mamayev and N.\,N. Trunov,
{Yadernaya Fizika} {\bf 37}, 1603 (1983);
{Sov. J. Nucl. Phys.} {\bf 37}, 952 (1983).

\bibitem{BMR98}
V.\,B. Bezerra, V.\,M. Mostepanenko, and C.~Ro\-mero,
{Int. J. Mod. Phys.} D {\bf 7}, 249 (1998).

\bibitem{BLMPv}
M. Bordag, J. Lindig, V.\,M. Mostepanenko, and Yu.\,V. Pavlov,
{Int. J. Mod. Phys.} D {\bf 6}, 449 (1997).

\bibitem{GrishchukY80}
L.\,P. Grishchuk and V.\,M. Yudin,
{J. Math. Phys.} {\bf 21}, 1168 (1980).

\bibitem{MTrunov03}
X.\,S. Mamaeva and N.\,N. Trunov,
{Teor. Mat. Fiz.} {\bf 135}, 82 (2003);
{Theor. Math. Phys.} {\bf 135}, 520 (2003).

\bibitem{Pv}
Yu.\,V. Pavlov,
{Teor. Mat. Fiz.} {\bf 126}, 115 (2001);
{Theor. Math. Phys.} {\bf 126}, 92 (2001).

\bibitem{Fulling79}
S.\,A. Fulling,
{Gen. Relativ. Gravit.} {\bf 10}, 807 (1979).

\bibitem{Pv4}
Yu.\,V. Pavlov,
{Teor. Mat. Fiz.} {\bf 140}, 241 (2004);
{Theor. Math. Phys.} {\bf 140}, 1095 (2004).

\bibitem{Lanczos38}
C. Lanczos,
{Ann. Math.} {\bf 39}, 842 (1938).

\bibitem{PvIJA}
Yu.\,V. Pavlov,
{Int. J. Mod. Phys.} A {\bf 17}, 1041 (2002).

\bibitem{NewtonWigner49}
T.\,D. Newton and E.\,P. Wigner, {Rev. Mod. Phys.} {\bf 21}, 400 (1949).

\bibitem{Schweber}
S. S. Schweber, {\it An Introduction to Relativistic Quantum Field Theory\/}
(Row, Peterson and Co, New York, 1961).

\end{thebibliography}
\end{document}